\RequirePackage[svgnames]{xcolor}

\documentclass{aa}  
\bibpunct{(}{)}{;}{a}{}{,} 
\definecolor{bluegrey}{RGB}{42, 54, 112}
\usepackage{graphicx}
\usepackage{siunitx}
\usepackage{natbib}
\defcitealias{dobbels20}{D20}
\DeclareSIUnit \parsec {pc}
\newcommand{\um}{\si{\micro\meter}}
\newcommand{\xum}[1]{\SI{#1}{\micro\meter}}

\usepackage{txfonts}

\usepackage[colorlinks=true, citecolor=bluegrey, linkcolor=bluegrey]{hyperref}
\begin{document}

\title{Predicting far-infrared maps of galaxies  \\
via machine learning techniques}

\author{Wouter Dobbels
	\and Maarten Baes
}

\institute{Sterrenkundig Observatorium, Universiteit Gent, Krijgslaan 281, B-9000 Gent, Belgium \\
	\email{maarten.baes@ugent.be}
}

\date{Accepted by A\&A, 24 August 2021}

\authorrunning{W. Dobbels \& M. Baes}
\titlerunning{Predicting FIR maps of galaxies via machine learning techniques}

\abstract
{The ultraviolet (UV) to sub-millimetre (submm) spectral energy distribution of galaxies can be roughly divided into two sections: the stellar emission (attenuated by dust) at UV to near-infrared wavelengths and dust emission at longer wavelengths. In \citet{dobbels20}, we show that these two sections are strongly related, and we can predict the global dust properties from the integrated UV to mid-infrared emission with the help of machine learning techniques.}
{We investigate if these machine learning techniques can also be extended to resolved scales. Our aim is to predict resolved maps of the specific dust luminosity, specific dust mass, and dust temperature starting from a set of surface brightness images from UV to mid-infrared wavelengths.}
{We used a selection of nearby galaxies retrieved from the DustPedia sample, in addition to M31 and M33. These were convolved and resampled to a range of pixel sizes, ranging from 150 pc to 3 kpc. We trained a random forest model which considers each pixel individually. }
{We find that the predictions work well on resolved scales, with the dust mass and temperature having a similar root mean square error as on global scales (0.32 dex and 3.15 K on 18\arcsec scales respectively), and the dust luminosity being noticeably better (0.11 dex). We find no significant dependence on the pixel scale. Predictions on individual galaxies can be biased, and we find that about two-thirds of the scatter can be attributed to scatter between galaxies (rather than within galaxies).}
{A machine learning approach can be used to create dust maps, with its resolution being only limited to the input bands, thus achieving a higher resolution than \textit{Herschel}. These dust maps can be used to improve global estimates of dust properties, they can lead to a better estimate of dust attenuation, and they can be used as a constraint on cosmological simulations that trace dust.}


\maketitle

%

\section{Introduction}

The radiation we detect from galaxies across wavelengths is the result of a complex interplay of different components within them, including stars, gas, and dust. Stars emit most of their energy between the ultraviolet (UV) and near-infrared (NIR). To a first order, the spectrum of stars can be described as a black body, with hotter (often short-lived) stars emitting at shorter wavelengths. Dust can absorb this stellar radiation, or scatter it into a different direction; the combination of these two effects is called attenuation. Since this attenuation grows stronger towards shorter wavelengths, the presence of dust leads to a reddening of the stellar spectrum \citep[see e.g. reviews by][]{galliano18, salim20}. Of course, more dust implies more attenuation, but the geometry of stars and dust is also of considerable importance \citep{witt92, witt00, baes01}. Young star clusters are often still enshrouded by dust from their birth clouds, and so the attenuation affects young stars more strongly than what is expected by a simple dust screen model.

The radiation from stars heats the dust to a typical value of 15 K to 25 K in the diffuse interstellar medium \citep{auld13, smith13, nersesian19}. A (modified) black body can again be used as a good first approximation, which implies that the dust emits in the far-infrared (FIR) to sub-millimetre (submm)  re\-gime. The large grains are in equilibrium with the local radiation field, and this leads to an energy balance: the total energy absorbed by dust must equal the total energy emitted. This is often used as a constraint in spectral energy distribution (SED) fitting models \citep{dacunha08, chevallard16, leja17, carnall18, boquien19}.

However, there is more that connects stars and dust than energy balance alone.  In a previous study \citep[][hereafter referred to as \citetalias{dobbels20}]{dobbels20}, we have shown that it is possible to estimate dust temperature from UV to mid-infrared (MIR) broadbands on a global scale. In addition, the dust luminosity and mass could be estimated from these broadbands, and these predictions are much better than what can be expected from energy balance alone. The accuracy of these predictions can be described by the root mean square error (RMSE). For the dust luminosity and mass, we found a RMSE of 0.16 and 0.30 dex, respectively, while the dust temperature could be estimated with a RMSE of 3.0 K. It is not surprising that there is a correlation between the stellar spectrum and dust properties, but this was the first study that thoroughly quantified this relation.

\citetalias{dobbels20} only investigated relations on a global scale, using the integrated broadband fluxes and properties derived from them. The current study is related to resolved scales, ranging from multiple kpc down to 150 pc. A priori it is unclear if the global relation is the result of very similar relations at resolved scales, or if the global relation emerges from a much wider variety of resolved relations. Energy balance, which is a robust approximation for sightlines of complete galaxies, no longer has to hold at resolved scales \citep{williams19}. The radiation that is absorbed in a pixel does not have to originate from that pixel, but it can come from other regions of the galaxy. This process can only be simulated with a full radiative transfer approach \citep{viaene17, viaene20, williams19, verstocken20, nersesian20, nersesian20b}.

The goal of this paper is to study the relation between stellar light, observed through a few UV to MIR broadbands, and dust, on a resolved scale. Combining all the broadbands and dust properties means we have to deal with a high-dimensional parameter space. Machine learning allows us to find an optimal, data-driven relation without having to limit ourselves to a subset of the parameters. The result is a model that can produce resolved maps of the dust properties, while only making use of UV to MIR images. The three dust properties that we consider are the (specific) dust luminosity, (specific) dust mass, and the dust temperature. The FIR SED can be fitted well with these three properties alone, in our case using the THEMIS dust model \citep{jones17}.  A careful uncertainty analysis allows us to also produce uncertainties on the predicted maps. This makes it possible to evaluate when and to what degree the predictions can be trusted.

Previous studies have investigated the use of machine learning models to predict galaxy properties from their SEDs \citep[e.g.][]{lovell19, surana20, simet21, gilda21}. One benefit is that these models can infer properties much faster than traditional SED fitting tools. In addition, they can be trained on mock observations of simulations, with the hope that this forward-modelling approach is less error-prone than a backward-modelling approach such as in SED fitting. In contrast, we focus on another use case: the input observations are limited in their wavelength range. In particular, SED fitting models fail to make an unbiased estimate of dust properties without the FIR \citep{dobbels20}. A machine learning approach can avoid this problem if it is trained on a set of galaxies that have FIR observations.

The resolution of our model is only limited by the resolution of the input bands. This means that we can reach even higher resolutions than that available from the \textit{Herschel} Space Observatory \citep{pilbratt10}. Although \textit{Herschel} has the largest mirror in space to date, its 36\arcsec resolution at \xum{500} is quite poor in comparison to optical telescopes which can reach sub-arcsecond resolutions. As we discuss in Sect.~\ref{ssec-highres-dustmap}, resolving a galaxy has multiple advantages. Even when only interested in global properties, they lead to improved estimates because they suffer less from the outshining bias \citep{galliano11, galametz12, utomo19}. The dust maps that we estimate can also be used to correct UV to NIR images for dust attenuation \citep{calzetti00, kennicutt09}. Finally, this relation between the UV--MIR and the dust properties is trained on observations, and can thus serve as a test for numerical simulations of galaxy evolution.

In the next section, we describe the data and methods that were used throughout this work. Sect.~\ref{sec-resultsML} shows the machine learning results for the three dust properties on different resolutions. In the same section, we also consider the predictions on individual galaxies, and show to which extent the predictions are reliable. Sect.~\ref{sec-interpretation} investigates the dependence on the pixel scale and discusses possible applications of our method. Finally, we conclude in Sect.~\ref{sec-conclusions}. In addition, appendix~\ref{app-additional-figures} contains additional figures which will be discussed in the main text.

\section{Data and methods}

\newcommand\Tstrut{\rule{0pt}{2.6ex}}       
\newcommand\Bstrut{\rule[-0.9ex]{0pt}{0pt}} 
\newcommand{\TBstrut}{\Tstrut\Bstrut} 
\begin{table}
	\caption{Summary of photometric broadbands used for this machine learning pipeline.  }
	\label{tab-bands}      
	\centering                                      
	\begin{tabular}{l c | l c}          
		\hline\hline                        
		Band  &  $\lambda_{\mathrm{piv}} (\um)$  & Band & $\lambda_{\mathrm{piv}} (\um)$  \TBstrut \\   
		\hline                                   
		\multicolumn{4}{l}{\textbf{Input bands}}\Tstrut \\
		GALEX FUV & 0.153 & 2MASS K$_{\mathrm{s}}$ & 2.16 \\
		GALEX NUV & 0.229 & WISE 1 & 3.37 \\
		SDSS u & 0.355 & WISE 2 & 4.62 \\
		SDSS g & 0.480 & WISE 3 & 12.1 \\
		SDSS r & 0.624 & WISE 4 & 22.2 \\
		SDSS i & 0.766 & IRAC 1 & 3.56 \\
		SDSS z & 0.908 & IRAC 2 & 4.50 \\
		2MASS J & 1.24 & IRAC 3 & 5.75 \\
		2MASS H & 1.65 & IRAC 4 & 7.92 \\
		\\
		\multicolumn{4}{l}{\textbf{Bands used to constrain dust properties (target)}} \\
		MIPS 1 & 23.6 & PACS \xum{160} & 161 \\
		MIPS 2 & 70.9 & SPIRE \xum{250} & 253 \\
		PACS \xum{70} & 70.7 & SPIRE \xum{350} & 354 \\
		PACS \xum{100} & 101 & SPIRE \xum{500} & 515 \\
		\hline                                             
	\end{tabular}
    \tablefoot{$\lambda_{\mathrm{piv}}$ represents the pivot wavelength of the band, calculated from the CIGALE transmission curves \citep{boquien19}.}
\end{table}

Our model is part of the domain of supervised machine learning. This means that we have a dataset with a corresponding ground truth. A subset of these data, the `training set', is then used to optimise a prediction model, in order for its predictions to be as close to the ground truth as possible. We can then validate our model on a separately kept `test set', to see how it performs on unseen data. For the purposes of this paper, the inputs of the model are the UV to MIR broadbands, from the GALEX \citep{martin05}, SDSS \citep{york00}, 2MASS \citep{skrutskie06}, WISE \citep{wright10} and \textit{Spitzer} \citep[IRAC only;][]{fazio04} telescopes. The ground truth consists of the dust properties, estimated from a full UV to FIR SED fit, which includes the previous telescopes as well as \textit{Herschel} \citep{pilbratt10} and the MIPS instrument on board \textit{Spitzer} \citep{rieke04}. The bands and their pivot wavelength are summarised in Table~\ref{tab-bands}, while their transmission curves are shown in Fig.~1 of \citet{clark18}.

The images of the different telescopes were convolved and resampled in order for the pixels to align spatially. We consider each spatial pixel as a separate sample, with a single flux per UV--MIR broadband. Each sample can then be identified by its galaxy name, pixel scale, and pixel position. However, we do ensure that galaxies are either fully in the training set or in the test set. In the next sections, we go into more detail about the used data sets and methods.

\subsection{Data}

In contrast to \citetalias{dobbels20}, the focus of this study is on resolved relations. To reach the highest possible resolution, we limit ourselves to nearby galaxies. In addition, we require UV to FIR observations of these galaxies, where the FIR is used to derive the ground truth. In \citetalias{dobbels20}, we used the aperture-matched photometry of the DustPedia dataset \citep{davies17}. By using the images of this dataset, we fulfil both our constraints of having nearby, UV-FIR images. 

The images were downloaded from the DustPedia archive\footnote{http://dustpedia.astro.noa.gr/}. The DustPedia photometry routine, named CAAPR and presented in \citet{clark18}, uses these images to produce aperture photometry fluxes.  We made a few small adjustments to this routine in order to output images instead of global fluxes. Most importantly, the extinction correction is applied to the image itself, instead of applying it to the aperture flux at a later step. We kept the same background subtraction routine. We limit the pixels to within $D_{25}$ (from the HyperLeda database; \citealp{makarov14}), since we found that the pixels outside $D_{25}$ were generally too noisy to be useful. This aperture is always smaller than the aperture that is used to extract the DustPedia fluxes.

We did, however, use a custom star subtraction routine. Instead of doing the subtraction (interpolation) immediately, we focussed on finding appropriate star apertures and masked the regions within. The astropy convolve routine can then properly handle these masked regions. For simplicity, we assume circular apertures. The star positions were taken from the GAIA DR2 database \citep{gaiacollaboration18}, where we used the parallax to ensure that the detected source lies within the Milky Way. We started by fitting 2D Gaussian profiles to the SDSS i band, at the GAIA positions and starting with a radius determined from the GAIA magnitude. We then iteratively fitted neighbouring bands, each time using the previous band as a soft constraint. The actual profile need not exactly match a Gaussian one: the fitted profile is only used to determine an appropriate radius of the star aperture. While most of the fitted apertures now seem fine upon visual inspection, we had to manually modify some of the apertures to prevent cutting out a significant part of the galaxy.

Two galaxies that were not included in the DustPedia set, due to their large angular extent, are M31 and M33. This study would, however, greatly benefit from these large, well-resolved galaxies. As such, we include these two galaxies as well. We used the UV to FIR data from \citet{viaene14} for M31 and from \citet{williams18} for M33. We reproduced the different steps in the CAAPR pipeline, from a polynomial sky subtraction to a \citet{schlafly11} based MW extinction correction. 

For DustPedia, we started from the same 715 galaxies as in \citetalias{dobbels20}, but further excluded 6 galaxies that had no SPIRE observations. After including M31 and M33, our data set contains 711 galaxies.

\subsection{Data preparation}

The images as described so far have been background subtracted, extinction corrected, and have circular masks at the star positions. However, each band can have a different pixel size. The pixel sizes range from 1.5" for GALEX to 12" for SPIRE \xum{500}. This means that the pixels of the different broadbands do not line up. Moreover, the different pixels are not independent. For example, the SPIRE bands are sampled at roughly one-third of the point-spread function (PSF) full width at half maximum (FWHM), with SPIRE \xum{500} having an approximate PSF FWHM of 36". While techniques exist to avoid being limited to the worst resolution \citep[e.g.][]{whitworth19}, we choose the simpler approach of convolving and resampling to the band with the worst resolution. The convolution kernels were taken from \citet{aniano11}, allowing us to convolve from any band to any SPIRE PSF. For the reprojection, we used the astropy affiliated `reproject' package\footnote{https://reproject.readthedocs.io}, using the `adaptive' method. When including SPIRE \xum{500}, this means that pixels are now of size 36", and the pixels across bands do align. Especially for the longer wavelength bands, the pixels are not fully independent (since the PSF extends beyond the FWHM), but this is a common approach \citep{bendo10, bendo12, viaene14, williams18} and our analysis does not require the pixels to be fully independent.

Besides including all bands and being limited to 36", we also redo this convolution and resampling to reach different pixel sizes. First of all, we excluded SPIRE \xum{500}, allowing us to reach a pixel size of 25" (the SPIRE \xum{350} PSF FWHM). Then, we excluded SPIRE \xum{350} and SPIRE \xum{500} to reach a pixel size of 18" (the SPIRE \xum{250} PSF FWHM). Furthermore, we selected a series of pixel sizes in physical units: 150 pc, 400 pc, 1 kpc, and 3 kpc. We convolved and resampled all galaxies that can reach a particular physical pixel size, if necessary by discarding SPIRE \xum{500} and/or SPIRE \xum{350}. Only ten galaxies (including M31 and M33) can reach the 150 pc scale. We also did not consider galaxies for which the semi-major axis is smaller than the pixel size (e.g. a galaxy with radius smaller than 3 kpc is not included for the 3 kpc scale).

The uncertainties on the pixel fluxes are calculated after the convolution, in a way similar to \citet{viaene14} and \citet{williams18}. We calculate the background variation uncertainty as the root mean square (RMS) of the background, after sigma clipping. The extinction correction uncertainty was calculated from the variation of the \citet{schlafly11} reddening maps. The calibration uncertainty was taken from various sources and is described in Table A.1 of  \citet{viaene14}. These different uncertainties are combined in quadrature.

A few steps are undertaken before feeding this data into a ML pipeline. These are the same steps as in \citetalias{dobbels20}, which are briefly summarised here. We fitted a Bayesian SED model through the UV to FIR fluxes of each pixel using the CIGALE code \citep{boquien19}. This gave us a Bayesian estimate of stellar and dust properties, with appropriate uncertainties. For the input, we made use of a UV-MIR CIGALE fit, and extracted a Bayesian estimate of the broadbands. This mitigates the problem of poorly sampled SEDs, where broadbands can be missing, negative, or vastly different from neighbouring bands. The Bayesian SED is a likelihood-weighted average of SEDs in a model library, described in \citet{nersesian19}.

Instead of feeding these Bayesian fluxes directly into the ML model, we first normalised all the fluxes by the WISE \xum{3.4} band. This ensures that the input is intensive, making it to first order independent of distance and number of stars in that pixel. Next, we took the logarithm of these normalised fluxes in order to reduce the large range of values. For the output, we also predict intensive properties: specific dust luminosity (s$L_d = L_d / L_*$), specific dust mass (s$M_d = M_d / M_*$), and dust temperature. For the first two, we always work in log space, in order to greatly reduce the dynamic range of these properties. 

\subsection{Machine learning}
\label{sect:method-ML}

As stated before, the ML model learns to predict an output by example of a training set. However, the goal must be to generalise well. It is of no use to perfectly predict every example of the training set, if the model does not provide accurate output for any example that is not part of the training set. When the model fails to generalise, we say it suffers from overfitting. There are multiple ways to deal with this, such as gathering more training data, using less complex models, and regularising the model parameters. 

We used a similar ML set-up as \citetalias{dobbels20}, with two important differences. First, we used a random forest instead of neural network, since those results were slightly better and more consistent. Second, we also included the log of the pixel scale in parsec to the input. This means that we can use all pixels as input, as opposed to training a separate model for each pixel scale. As in \citetalias{dobbels20}, we used a 4-fold train test split. This means that we divide our galaxy sample randomly in four equally sized parts, and then train four separate models, each tested on a different fold (25\% of the data), and trained on the remaining three (75\% of the data). Each of the four models thus has access to 533 training galaxies and 178 test galaxies. The number of pixels varies between the different train-test splits (especially considering that M31 has 61\,290 pixels at this scale), with a median of 120\,794 pixels for training.

We note that this method considers pixels individually. There exist powerful deep learning tools, such as convolutional neural networks, that allow using the complete image of a galaxy as input and produce a predicted image as output. In theory, these can use spatial information (such as the presence of dust lanes) to assist the predictions \citep{dobbels19}. However, using images as a whole, we only have 715 data points. Most image to image methods use thousands, if not hundreds of thousands of training images \citep{isola17}. By using the pixels individually, we have more than 300\,000 data points to train on, more than enough to avoid heavy overfitting. We have tried to add neighbouring pixels to the input as well, but this did not improve the results. Adding the radius from the centre as an additional input also did not improve results.

We follow \citetalias{dobbels20} in the prediction of aleatoric uncertainty. This is the uncertainty that is the result of both measurement uncertainties (on input and ground truth) and the lack of information that the input gives about the output. The UV--MIR gives some information to predict dust properties, but not enough information for perfect predictions. We again used a different model, since we found it to be somewhat better and more consistent. The model used is a gradient boosting machine, implemented using xgboost \citep{chen16}. This is a tree-based algorithm which allows for a self-defined loss function. The uncertainty is predicted by minimising the negative log-likelihood assuming Gaussian uncertainties (see \citetalias{dobbels20}).

\section{Results: Machine learning}
\label{sec-resultsML}

Throughout this section we compare the predicted values to the ground truth. These predictions are of course only meaningful when the ML model is evaluated on a test set, which has been kept separate from the training set. As stated in the previous section, we divide our galaxies in four folds, and train four models each tested on a different fold. The results shown combine the test folds of all four models.

\subsection{Dust properties at a fixed scale}

\begin{figure*}
	\centering
	\includegraphics[width=17cm]{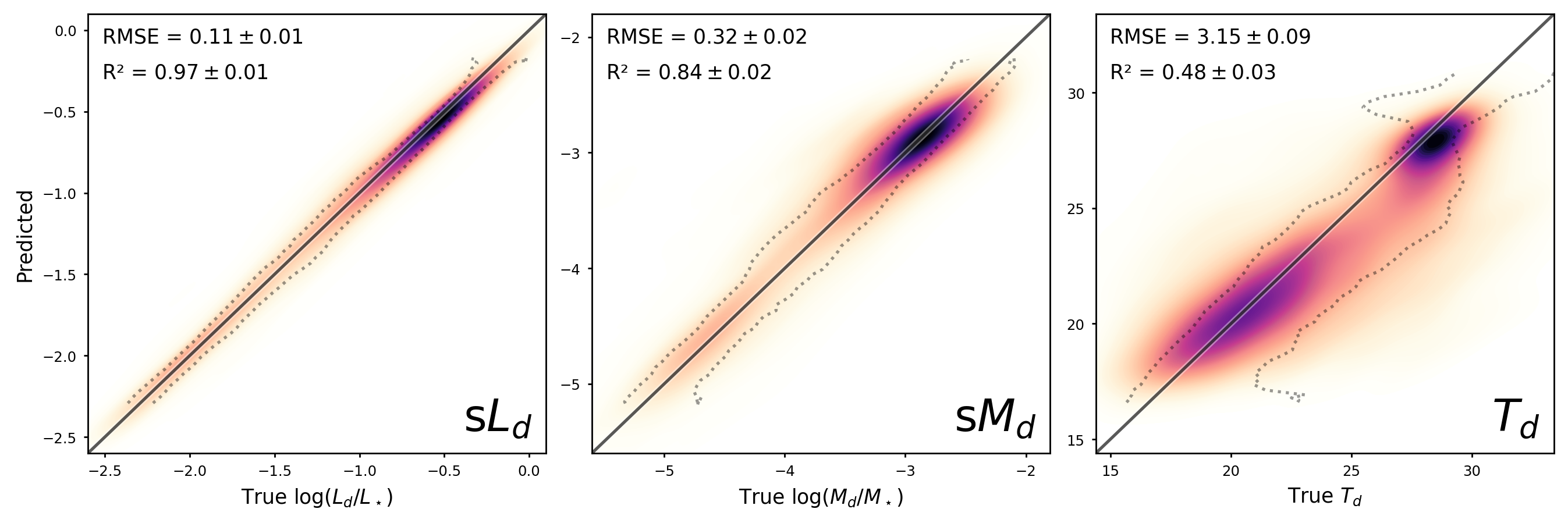}
	\caption{Comparing predicted to ground truth specific dust luminosity, specific dust mass, and dust temperature, on pixel scales of 18\arcsec . The colours are based on a 2D kernel density estimate (KDE), where each pixel is weighted inversely proportional to the number of pixels in that galaxy. The one to one line is denoted by a solid black line. The dotted black lines are the 16\textsuperscript{th} and 64\textsuperscript{th} percentiles after binning the predicted values. }
	\label{fig-truevspred-18arcsec}
\end{figure*}

We are interested in three different dust properties: s$L_d$, s$M_d$, and $T_d$. To summarise the predictions of these properties, we start by considering only the (angular) pixel scale at 18\arcsec . Fig.~\ref{fig-truevspred-18arcsec} compares the predictions to the ground truth, by using a kernel-density estimate (KDE) to colour the plot. The pixels are assigned a weight, inversely proportional to the number of pixels in that galaxy. This ensures that each galaxy, not each pixel, contributes equally. In other words, this avoids large galaxies (like M31) from dominating the results, since all its pixels will be assigned a lower weight. These large galaxies will be investigated further. The weighting influences the kernel density estimate, as well as the prediction scores given in the top left of each panel. These metrics also include an uncertainty estimate, derived from bootstrap sampling the galaxies.

Consistent with \citetalias{dobbels20}, we find the best predictions for s$L_d$, followed by s$M_d$, and finally by $T_d$. The RMSE scores are also similar to those on global scales in \citetalias{dobbels20}. For s$L_d$, we find a RMSE of  0.11 dex (0.16 dex on global scales), with all the predictions falling very close to the ground truth. Surprisingly, these predictions are better than on global scales. This is despite the energy balance approximation, which is more accurate at larger scales. \citet{williams19} find that due to the effects of non-local heating, energy balance breaks down at scales smaller than about 1.5 kpc. As described in \citetalias{dobbels20}, the MIR is useful to predict s$L_d$, and this tracer does not rely on energy balance. When excluding WISE, we find a RMSE of 0.17~dex (0.30~dex on global scales), showing indeed that the MIR is an important but not required contribution for s$L_d$ predictions. Since we consider pixels as individual data points, resolving the galaxies leads to a much larger dataset; this can also help explain why these predictions are somewhat better than on global scales.

A variable that is strongly linked to both the UV--MIR fluxes as well as s$L_d$, is the star-formation rate (SFR). This parameter can be estimated from fitting the stellar SED \citep[e.g.][]{salim07, maraston10, conroy13}, as well as from the total dust luminosity \citep[e.g.][]{bell03, calzetti07, elbaz11}. Young stars are still enshrouded by their birth clouds, and much of their radiation is absorbed (and reradiated) by dust; some of the emission escapes, leaving an important signature in the UV \citep{kennicutt98a, kennicutt12}. Since the SFR is strongly linked to both our input (UV--MIR fluxes) as well as our prediction target (s$L_d$), this can help explain why our predictions work so well, even on resolved scales.

The specific dust mass has a RMSE of 0.32 dex, with somewhat more scatter than s$L_d$ but no clear bias. On global scales we found an RMSE of 0.30 dex, with the difference between the two not being statistically significant. The dust temperature is less accurate, with a RMSE of 3.15 K, and again no significant difference with the RMSE of 3.00 K found on global scales. Of course the RMSE, with units K, can not be directly compared to that of s$L_d$ and s$M_d$, but the dimensionless $R^2$ score (coefficient of determination, also known as the fraction of explained variance) can. While about 96\% of the variance in s$L_d$ can be explained, and 83\% of the variance in s$M_d$, only about half of the variance in $T_d$ can be predicted from the UV-MIR input. We see a clear bimodality in dust temperature, and we can make quite a good distinction between the two modes, but have inaccurate estimates within that mode. This bimodality can also be seen in \citet{nersesian19}, with early-type and irregular galaxies having high dust temperatures (around 30~K), and late-type galaxies having lower dust temperatures (around 20~K). Although the dust temperature can not be predicted as well as the other two properties, the predictions are much better than random, and this means that we can go beyond what is possible from energy balance assumptions or MIR scaling relations alone.

\subsection{Dust properties at various scales}
\label{sect-crossscale}

While the previous section indicates that the results of \citetalias{dobbels20} extend well to resolved scales, we do have to keep in mind that the majority of our sample (67\%) has a distance between 10 Mpc and 30 Mpc. When using the 18\arcsec pixels as before, this translates to pixel sizes from 0.9 kpc to 2.6 kpc, which, compared to the median D$_{25}$ of 13.5 kpc, is still quite large. In this section, we investigate the quality of the predictions at various scales, down to 150 pc.

\begin{figure*}
	\centering
	\includegraphics[width=17cm]{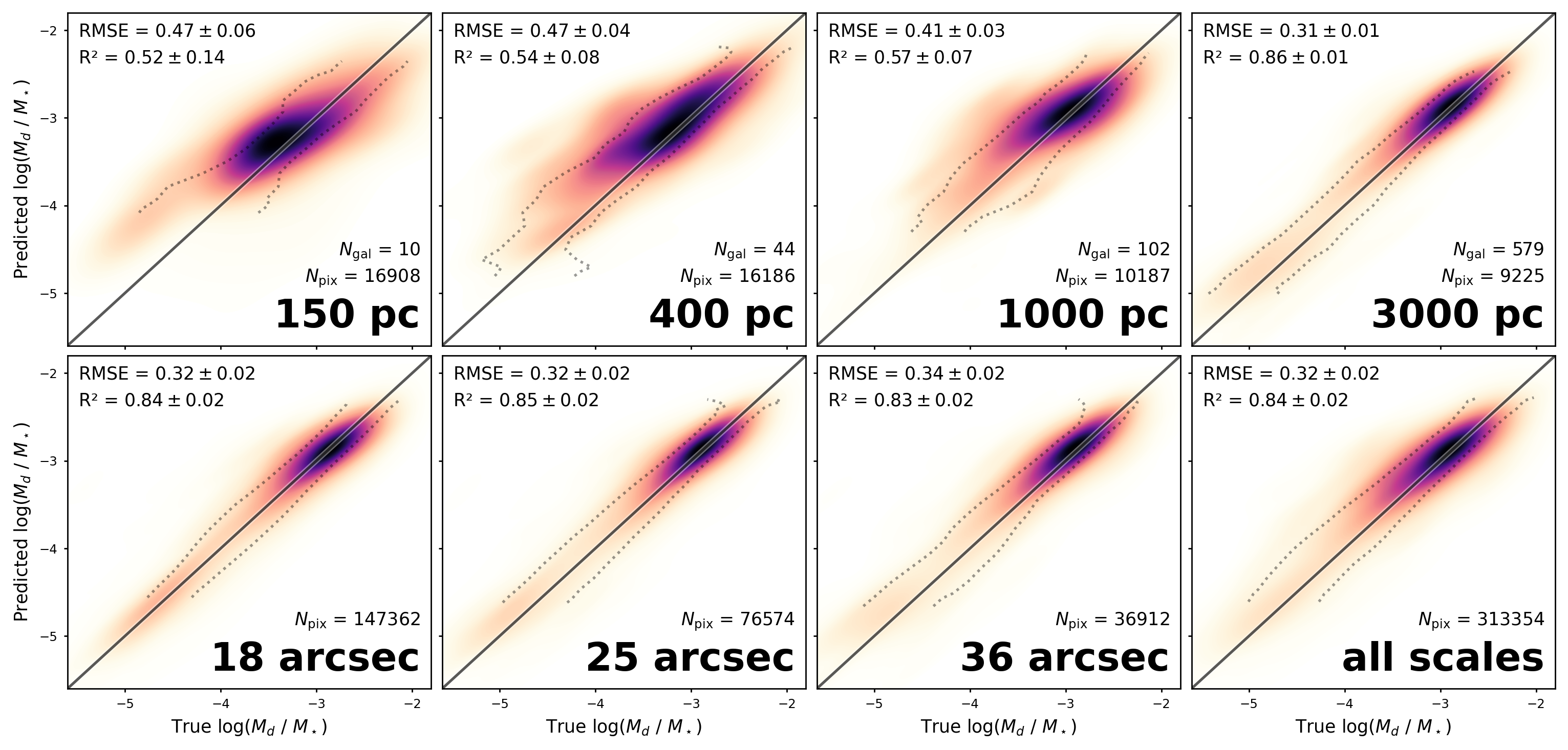}
	\caption{Similar to Fig.~\ref{fig-truevspred-18arcsec}, but showing only the specific dust mass at various (physical and angular) pixel scales. The number of galaxies and number of pixels (total, combining all four test sets) is indicated at the bottom right of each panel. The bottom right panel combines all pixels of the other panels.}
	\label{fig-truevspred-sMd}
\end{figure*}

In Fig.~\ref{fig-truevspred-sMd}, the predictions of s$M_d$ are compared to the ground truth. The bottom left panel is the same as the middle panel of Fig.~\ref{fig-truevspred-18arcsec}. The top row discards further away galaxies in order to reach smaller physical scales. On first sight, this row indicates that smaller scales have worse predictions than larger scales. However, we have to be careful because the samples in each of these panels are different. More specifically, smaller pixel scales mean more nearby and faint galaxies, including dwarfs; for our sample we indeed find that the 150 pc and 300 pc scales have a large share of low luminosity galaxies. To avoid comparing different samples, we investigate the 27 galaxies that can reach scales from 400 pc to 3 kpc. Calculating the RMSE for this shared set of galaxies, we find a RMSE of $0.39 \pm 0.04$ dex at 400 pc, $0.40 \pm 0.04$ at 1 kpc, and $0.36 \pm 0.04$ at 3 kpc. As such, we can not conclude that there is a trend between pixel size and prediction accuracy.

The predictions of dust luminosity and temperature at various scales are shown in Appendix~\ref{app-additional-figures}. We see that for both of these properties, the results also degrade towards smaller scales (excluding 150 pc) when including all galaxies. Still, the dust luminosity remains very accurate over all scales. When using the 27 galaxies from 400 pc to 3 kpc, we find no trend between pixel size and prediction accuracy. At 400 pc and 150 pc, the dust temperature also shows some bias from the one-to-one relation, likely due to the small number of galaxies. As we see later, the relation between UV-MIR and dust temperature can vary strongly between galaxies.

\subsection{Per-galaxy residual}
\label{sec-globalbias}

The scatter between the predictions and ground truth can be seen as a combination of scatter within each galaxy, and scatter between galaxy-averaged values. Because pixels within a galaxy will often be strongly spatially correlated, the latter term can be significant. To measure this, we compare the average of all predictions of a galaxy, to the average of the ground truth, revealing a global bias of that galaxy. The RMSE of these galaxy-averaged values will always be lower than or equal to the overall RMSE (which as discussed before, weights each pixel in order for all galaxies to have equal weight). On one extreme, the galaxy-averaged predictions would always match the average ground truth, resulting in the overall RMSE being solely from within-galaxy scatter. The other extreme arises when the residual of each pixel is equal to the residual of the galaxy average. In this case, the within-galaxy scatter disappears, and the overall RMSE only stems from the scatter between galaxies. 

\begin{figure*}
	\centering
	\includegraphics[width=17cm]{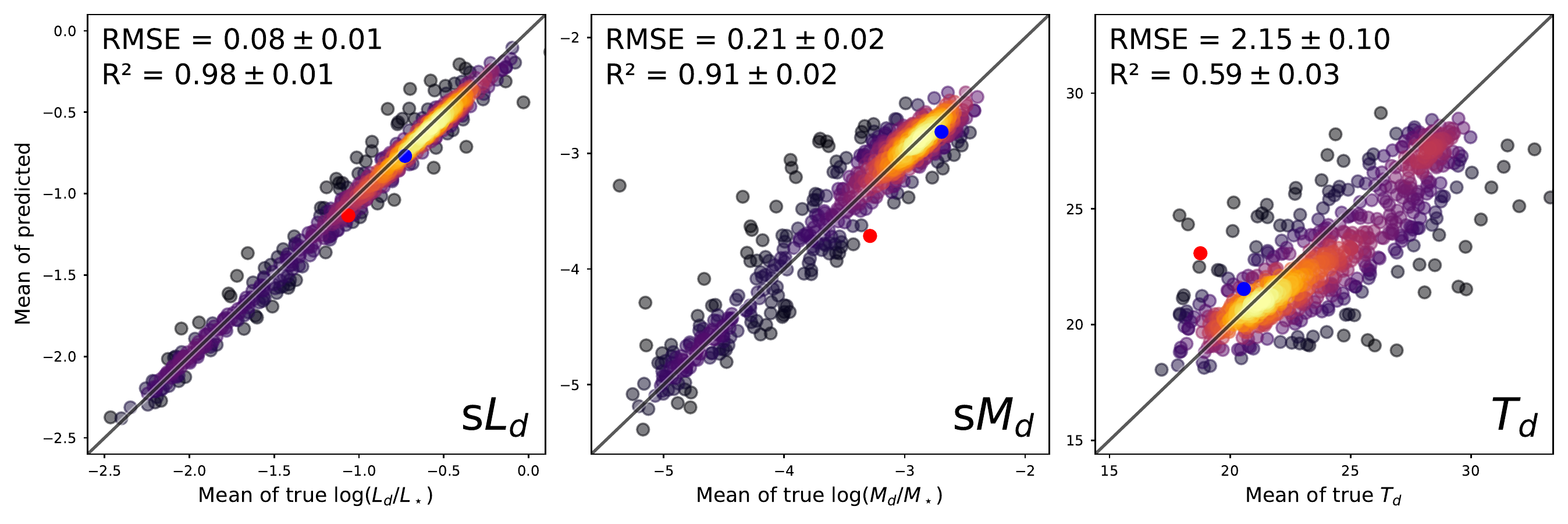}
	\caption{Similar to Fig.~\ref{fig-truevspred-18arcsec}, but averaged over all pixels per galaxy. Each dot shows a galaxy, coloured by a KDE. Highlighted are M31 in red and M33 in blue. The images from which the averages are taken have a resolution of 18\arcsec. When taking the mean of a property, all pixels within a galaxy are weighted equally.}
	\label{fig-globalbias}
\end{figure*}

The galaxy-averaged results for the three considered dust properties are shown in Fig.~\ref{fig-globalbias}. The RMSE for all three properties is about 68\% of the overall RMSE (68\% for both s$L_d$ and $T_d$, 65\% for s$M_d$). This shows that indeed a majority of the scatter can be explained as scatter between different galaxies. M31 and M33 have a well predicted s$L_d$, but for M31 the dust mass is strongly underpredicted and its dust temperature is strongly overpredicted. We further investigate these two galaxies in Sect.~\ref{sec-individual-galaxies}.

This between-galaxy scatter, which results in predictions which are biased for individual galaxies, is not surprising. After all, different galaxies can have different dust properties, star--dust geometries, and dust heating mechanisms \citep{bendo15, clark19, verstocken20, nersesian20}. The UV--MIR information of a single pixel is not enough to determine all these factors. We have experimented with adding additional information, such as the Hubble stage of the galaxy or the distance to the centre of that galaxy, as input to the pipeline. These somewhat improve the predictions on smaller scales for some galaxies, but this does not avoid all the bias. Maybe more complex morphological information, such as the convolutional neural networks, could be of use. However, the sample that we use in this work is much smaller, and thus adding a large number of inputs will likely lead to overfitting.

While the between-galaxy scatter is indeed in the majority, the within-galaxy scatter is not negligible: the pixels in a galaxy are not perfectly correlated. An interesting consequence of this is that the images can be used to improve the predictions on global scales. For example, \citetalias{dobbels20} predicted the global dust mass, and found a RMSE of 0.30 dex. If instead we use the images of 18\arcsec, we can sum the predictions to also get a predicted global dust mass. Doing this, we find a lower RMSE of 0.24 dex. This RMSE is not equal to the 0.21~dex reported in Fig.~\ref{fig-globalbias}: the figure reports an unweighted average of the logarithmic dust mass over all pixels in a galaxy, while for the global dust mass the more massive pixels will contribute more. Even if one would only be interested in global properties, the images provide extra information that results in an improved prediction. We come back to this in Sect.~\ref{ssec-highres-dustmap}.

\subsection{Predicted uncertainties}

While predicting the dust properties is useful, these predicted values are meaningless unless there is a corresponding uncertainty estimate. The simplest approach is to use a constant uncertainty, derived from the results on the test set. For example, the specific dust mass on 18\arcsec scales has a RMSE of 0.32 dex on the test set, and this can be seen as the expected uncertainty on new predictions.

\begin{figure}
	\resizebox{\hsize}{!}{\includegraphics{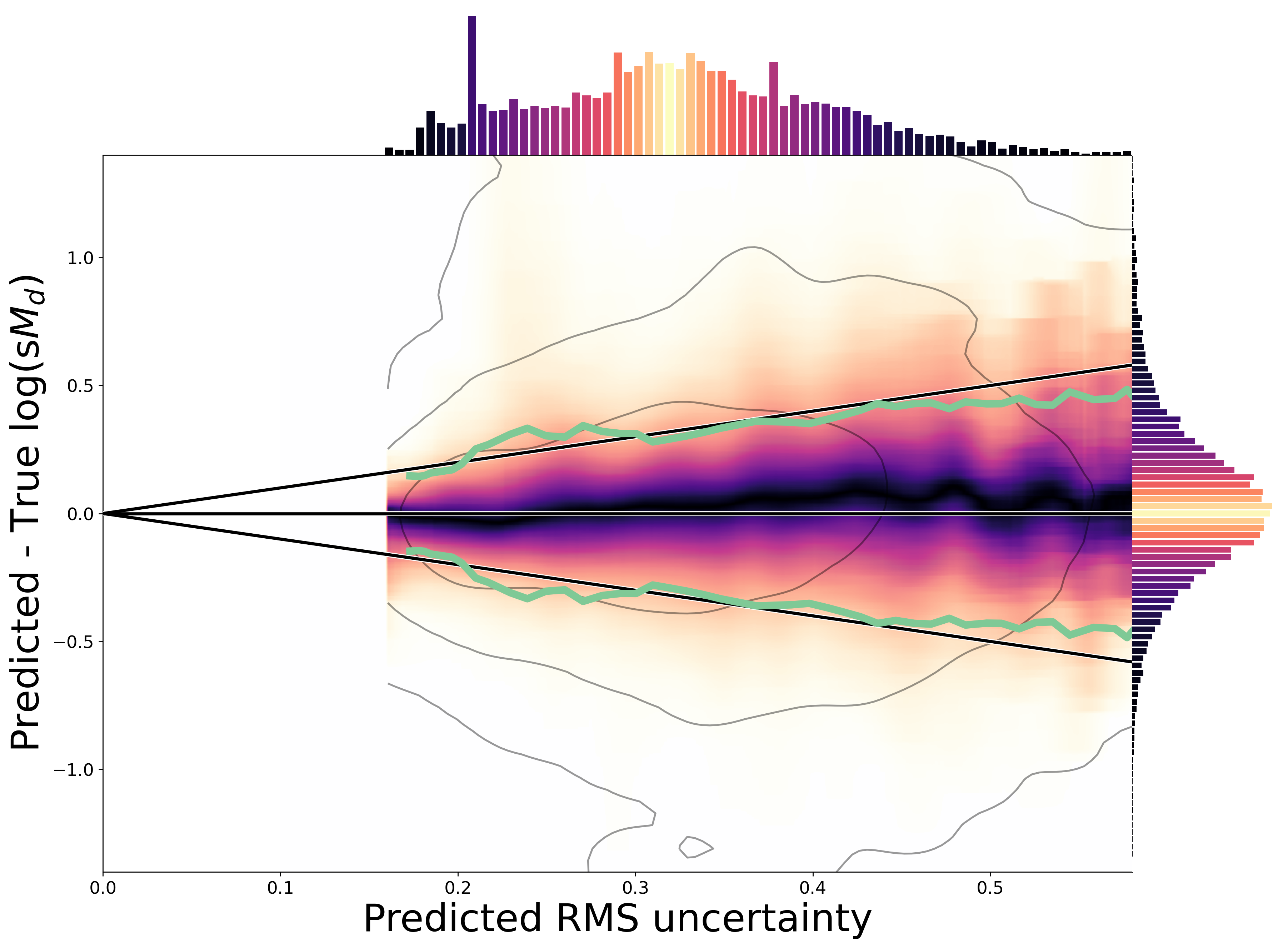}}
	\caption{Validating the predicted uncertainties of the specific dust mass on 18\arcsec scales. The ordinate shows the difference between the predicted log(s$M_d$) and the ground truth, also known as the prediction error. The abscissa shows the uncertainty predicted by the xgboost model. The colour is not a KDE estimate as in the previous figures, but instead shows the percentile, with the brightest values being used for the median. The bright green line shows a binned RMS of the prediction error. Both the percentile and RMS are binned in the abscissa, i.e. always considering a constant predicted uncertainty. The contours show a KDE estimate and these contours contain 68.3\%, 95.5\% and 99.7\% of the pixels. As done before, all pixels are weighted inversely proportional to the number of pixels in that galaxy.}
	\label{fig-uncval}
\end{figure}

While this approach is valid, we can go further. As described in Sect.~\ref{sect:method-ML}, we use a second model to predict uncertainties on individual pixels. This predicted uncertainty should match the root mean square (RMS) of the prediction error (the residual of the regression model). This is confirmed in Fig.~\ref{fig-uncval}, which shows the s$M_d$ on a log scale, at 18\arcsec. The green line, showing the binned RMS of the prediction error, matches the black one-to-one lines very well. The lowest predicted uncertainty is at 0.16 dex. A few pixels extend to predicted uncertainties above 1, but most of the values are below 0.5 dex. In each bin of predicted uncertainty, the prediction error is approximately Gaussian, and the standard deviation matches well with the predicted uncertainty.

To quantify the uncertainty prediction, we use the $\chi^2$ as defined in \citetalias{dobbels20}, except that we weight the pixels as before. In order to correct the uncertainty predictions, we set apart a validation set (split from the training set), as done in \citetalias{dobbels20}. When testing the uncertainty predictions on the test set, we find an overall $\chi^2$ for s$M_d$ of 1.04. Since ideally this metric should be close to one, we can conclude that the uncertainty predictions work well. However, the $\chi^2$ values can vary strongly between galaxies, and this distribution is skewed: for the mean $\chi^2$ of individual galaxies, we find a 16\textsuperscript{th}, median, and 84\textsuperscript{th} percentile of 0.30, 0.60, and 1.33. M31 and M33 have a mean $\chi^2$ of 4.04 and 0.85 respectively. With the majority of the prediction error being attributed to variation between galaxies (as discussed in the previous section), it is not surprising to see this variation in $\chi^2$ between galaxies. The global $\chi^2$ from \citetalias{dobbels20} can be used to estimate the prediction bias per galaxy, after which the per-pixel $\chi^2$ becomes a more useful measure of the uncertainty of pixels within each galaxy.

\subsection{Individual galaxy predictions}
\label{sec-individual-galaxies}

\begin{figure*}
	\centering
	\includegraphics[width=17cm]{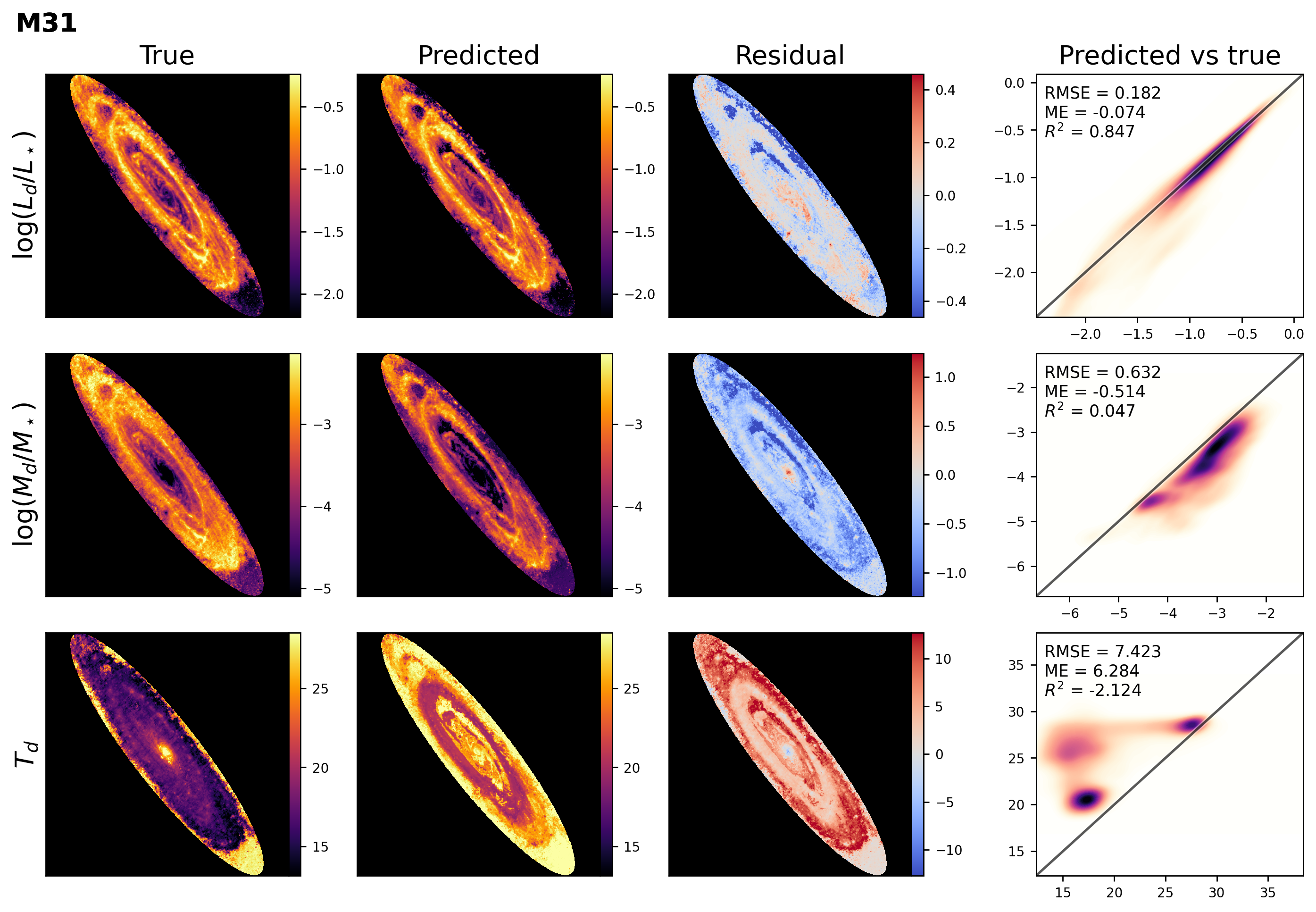}
	\caption{Overview of the predictions on M31 at an 18\arcsec scale. The first three columns show images of the ground truth, ML predictions, and residual (prediction - truth) respectively. The final column compares the ground truth and prediction, similar to Fig.~\ref{fig-truevspred-18arcsec}. The rows represent s$L_d$, s$M_d$ and $T_d$, respectively. In addition to the RMSE and $R^2$ scores, the mean error (ME) is also included in the top left.}
	\label{fig-summary-M31}
\end{figure*}

\begin{figure*}
	\centering
	\includegraphics[width=17cm]{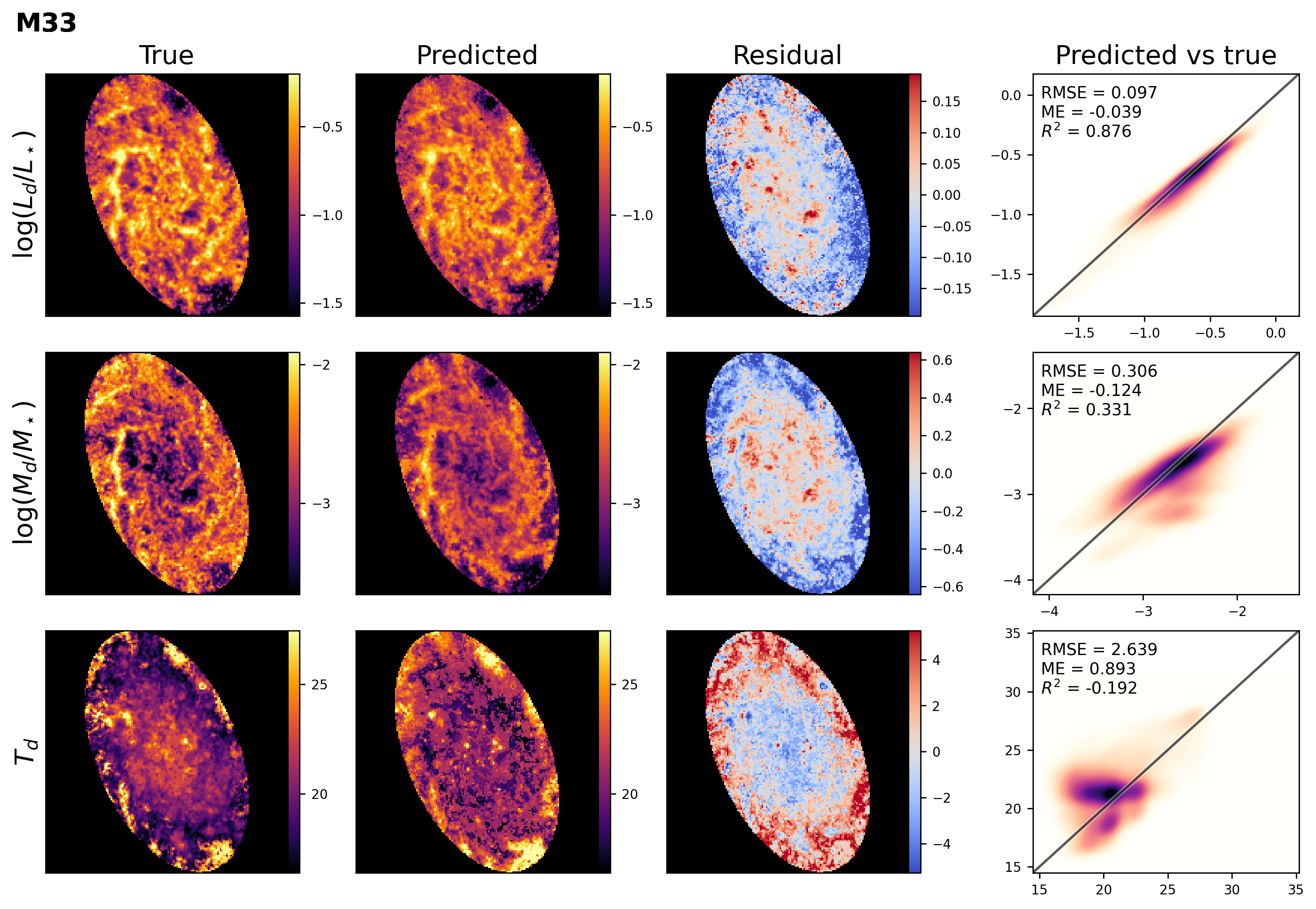}
	\caption{Similar to Fig.~\ref{fig-summary-M31}, but for M33.}
	\label{fig-summary-M33}
\end{figure*}

So far, we have investigated the ML predictions on the overall sample. We now turn to the predictions on individual galaxies. We limit ourselves to the most large and nearby galaxies in the sample, starting with the two large galaxies in our local group. An overview of the M31 predictions is shown in Fig.~\ref{fig-summary-M31}, while the M33 overview can be found in Fig.~\ref{fig-summary-M33}. 

For both of these nearby galaxies, s$L_d$ is predicted very accurately, with little bias. We see that especially the brighter regions, such as the dust ring of M31, are recovered very well. Some bias can be present in fainter regions, especially towards the outskirts, although this might be (partially) attributed to noise. The residual shows some structure, but this can be expected since particular regions in a galaxy will be highly correlated. 

As already noted in Sect.~\ref{sec-globalbias}, we find that for M31 the specific dust mass is underpredicted, while the dust temperature is overpredicted. These two biases are connected, since the three dust properties are not independent. For a constant dust mass, warmer dust will be more emissive. Again we notice that the bright dust ring of M31 is reproduced quite well. The centre is also reproduced well, it is the only region that does not systematically underpredict s$M_d$ and overpredict $T_d$. The interarm regions, in contrast, show large biases. 

This bias for M31 is not entirely unexpected: after all, M31 is not a typical spiral galaxy. \citet{viaene17} find that about 91\% of the dust in Andromeda is heated by old stars. This is mostly due to the bright bulge, which can heat the spiral arms even up to the outer disk. Since our pipeline considers each pixel individually, it cannot detect this non-local heating. When we inspected the worst predictions in the interarm regions, we find that these pixels are much colder than similar pixels in other galaxies. In \citetalias{dobbels20}, we found that the WISE \xum{22} - WISE \xum{4.6} had one of the largest correlations with the dust temperature. The relation between this MIR colour and the dust temperature shows a bimodality, and these interarm regions have a MIR colour in the transition region between the two modes. Unfortunately, our model predicts these fluxes to be in the high temperature mode, while in reality they seem to be part of the low temperature mode.

M33 shows less bias, although we can again identify spatial structures in the residuals. Towards the centre, we mostly find patches of overestimated s$L_d$ and s$M_d$, with a dust temperature that tends to be underestimated. The opposite happens towards larger radii, where we tend to overestimate $T_d$ but underestimate s$L_d$ and s$M_d$. The dust temperature, which almost exclusively falls in the low temperature mode of Fig.~\ref{fig-truevspred-18arcsec}, has a below average RMSE of 2.66 K. However, we do not seem to recover the spatial structure, in particular the decreasing temperature with increasing radius. In contrast to M31, M33's dust heating is dominated by heating of young stars \citep{williams19}.

\begin{figure*}
	\centering
	\includegraphics[width=17cm]{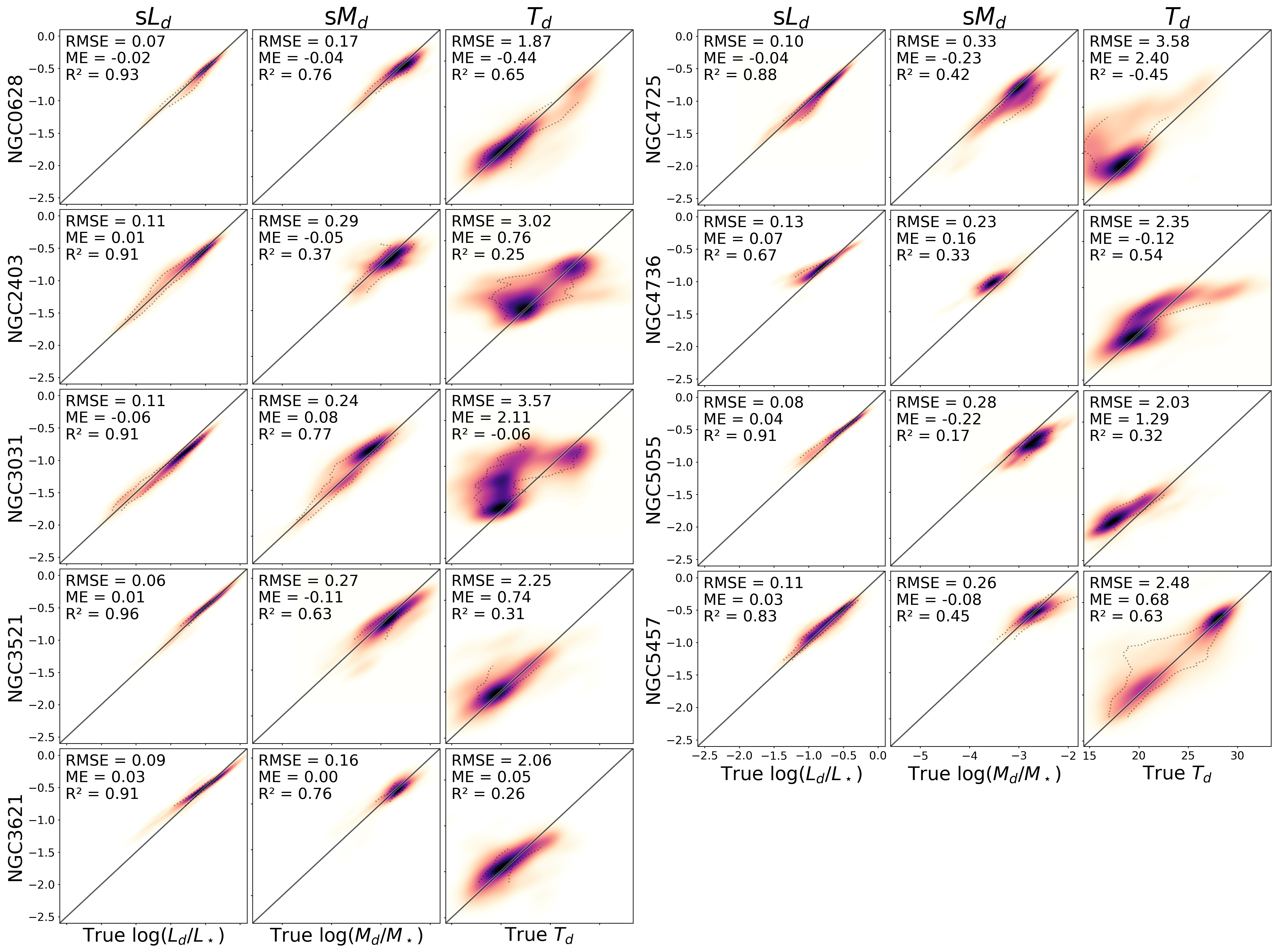}
	\caption{Similar to Fig.~\ref{fig-truevspred-18arcsec}, but for the nine nearby DustPedia galaxies in the sample of \citet{baes20}. }
	\label{fig-summary-nearby}
\end{figure*}

Next, we broaden our scope and look at nine nearby DustPedia galaxies from the sample of \citet{baes20}. The predictions are compared to the ground truth in Fig.~\ref{fig-summary-nearby}. When measured by RMSE, the quality of these predictions is slightly better than the overall sample: the RMSE of s$L_d$ is 0.098 dex (0.111 dex for the overall sample), for s$M_d$ it is 0.255 dex (overall 0.324 dex) and for $T_d$ it is 2.654 K (overall 3.152 K). This might be attributed to these nearby galaxies having high quality observations (which leads to both a more accurate input as well as a better ground truth). 

Whereas the overall sample shows little to no bias, all three properties can have biased predictions for individual galaxies. For example, the specific dust luminosity for NGC3031 is on average 0.058 dex higher than predicted. However, this bias tends to be small compared to the overall data variance. While for some the dust temperature remains problematic, most of these nearby galaxies show a remarkably well-predicted dust temperature.

\section{Interpretation and discussion}
\label{sec-interpretation}

In \citetalias{dobbels20} we have already investigated what drives the relation between the UV--MIR fluxes and the FIR. There we concluded that the MIR bands are useful to predict the \xum{70} and \xum{100} fluxes. The predictions for longer wavelengths mainly rely on observations around \xum{1}. In this paper, we want to investigate whether the relation between UV--MIR and the FIR is scale dependent. Equivalently, we investigate if the relation between the (dust attenuated) stellar SED and the dust properties is scale dependent. 

\subsection{Dependence on input pixel scale}

\begin{figure*}
	\centering
	\includegraphics[width=17cm]{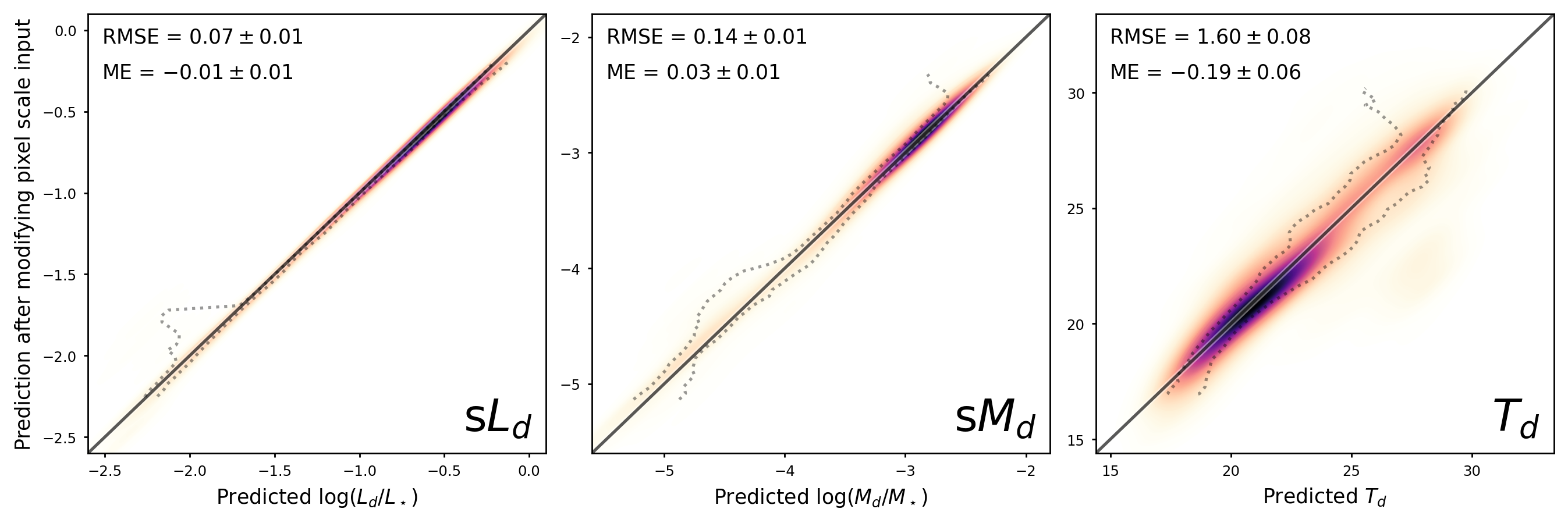}
	\caption{Comparing the predictions of the fiducial model on pixels of size 3 kpc to predictions on the same data, but after modifying the input pixel scale to 400 pc.}
	\label{fig-modify-scale}
\end{figure*}

Our machine learning model captures the relation between the stellar SED and the dust properties. Its input consists of the colours with WISE \xum{3.4}, the logarithm of the WISE \xum{3.4} intensity, and the logarithm of the pixel scale in parsec. This last input can be used to probe the scale dependence of the star--dust relation. We start from our fiducial models for each of the three dust properties. We make a prediction for each of the properties for all pixels on a 3 kpc scale. Next, we modify the input pixel scale to 400 pc, and make predictions for the same dataset. Essentially, this means we lie to the model about the actual pixel scale: the input is still derived from the images at a 3 kpc scale, but we tell the model they are at a 400 pc scale. In Fig.~\ref{fig-modify-scale}, we compare these two predicted values to each other.

It is clear that the pixel size has little impact on the predictions. The relation between UV--MIR and the dust properties seems to be the same across pixel scales. This is especially true for the specific dust luminosity. The specific dust mass and the dust temperature show some scatter. This means that for some pixels, the 400 pc mass or temperature can be a bit higher than expected at 3 kpc, and for other pixels it can be a bit lower. A small bias is also present: at 400 pc the dust mass is slightly higher and the dust temperature is slightly lower than expected at 3 kpc. According to the bootstrap uncertainty on the mean error metric, these biases are significant at a 3$\sigma$ level. Still, it is clear that this bias is small compared to the scatter, especially considering that there is also an additional scatter between the unmodified predictions and the ground truth.

These results make use of the fiducial model, and modify the pixel scale at inference without retraining the model. An alternative way to check the dependence on the pixel scale is to train a model on one scale, and test it on another. We found that these models worked well, showing again that pixel scale is not an important input. As an extreme case, we trained a model only using the global fluxes, and tested this model on 400 pc scales. For the specific dust luminosity, this model achieves a (weighted) RMSE of 0.23 dex (compared to 0.15 dex for the fiducial model; see Fig.~\ref{fig-truevspred-sLd}), with no significant bias. For specific dust mass, the model's RMSE is 0.51 dex (compared to 0.47 dex as shown in Fig.~\ref{fig-truevspred-sMd}), and no significant bias (ME = $0.04 \pm 0.05$). Finally for the dust temperature, the RMSE is 5.00 K (compared to 4.67 K in Fig.~\ref{fig-truevspred-Td}), with some bias to underpredict (ME = $-0.88 \pm 0.51$ K). While these predictions are worse than our fiducial model, the training did not require resolved images at different pixel scales. Considering that, we find that the difference with our fiducial model is quite small, and thus we again confirm that there is little dependence on pixel scale, at least down to 400 pc.

On first sight, this might contradict the results of \citet{williams19}. They found that energy balance breaks down at scales below 1.5 kpc. However, energy balance still holds \textit{on average}: some pixels will have excess FIR emission (compared to their attenuation), while other pixels will have a deficit. Without additional knowledge from neighbouring pixels, our model can not make this distinction, and thus its prediction is not altered by changing the input scale. 

\subsection{Creating high resolution dust maps}
\label{ssec-highres-dustmap}

\begin{figure*}
	\centering
	\includegraphics[width=17cm]{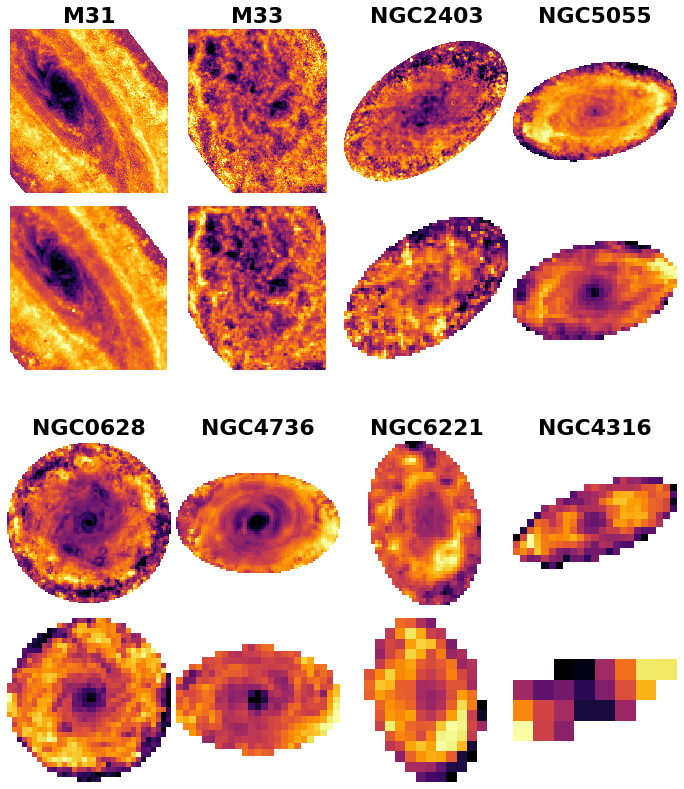}
	\caption{Comparing the predicted specific dust mass (top), at a pixel size of 6.7\arcsec, to the ground truth specific dust mass (bottom) at a pixel size of 18.4\arcsec. A selection of eight galaxies is made, and these are ordered by angular size.  M31 and M33 are zoomed in to reveal more details.}
	\label{fig-hires-predictions-sMd}
\end{figure*}

As discussed in the introduction of this paper, one of the limitations of the far-infrared is the poor angular resolution of the current generation of FIR missions. When comparing our predictions to a ground truth, we are also limited by this resolution. However, when applying our model we are only limited by the resolution of the worst input band, in our case WISE \xum{22}. While the SPIRE \xum{250} PSF FWHM is about 18\arcsec , the WISE \xum{22} PSF FWHM is about 12\arcsec \citep{aniano11}. By limiting the range of input bands to shorter wavelengths, even higher resolutions can be achieved at the cost of less accurate predictions. In the previous section we showed that the model is almost independent of pixel size. This gives us faith that the model can be applied at these higher resolutions to predict dust properties, while retaining the accuracy shown in Fig.~\ref{fig-truevspred-18arcsec} and Fig.~\ref{fig-truevspred-sMd}. Of course, we can not verify this by comparing to the ground truth, since the ground truth is limited by the SPIRE resolution.

Instead, we can use the (lower resolution) ground truth to improve the high resolution predictions, by requiring that the resampled prediction matches the ground truth. While this high resolution prediction is not an observation, it is an inference based on correlations of observed galaxies. Such predictions for the specific dust mass on a selection of eight galaxies is shown in Fig.~\ref{fig-hires-predictions-sMd}. The top rows show the predictions on the 6.7\arcsec resolution of the WISE \xum{12} band, which means that the input was also limited to wavelengths up to this band. The row below shows the ground truth at the 18.4\arcsec resolution of the SPIRE \xum{250} band. The higher resolution reveals a range of prominent features, such as more clearly defined spiral arms and dust rings. Complementary figures showing the predictions for s$L_d$ and $T_d$ are shown in appendix~\ref{app-additional-figures}: Fig.~\ref{fig-hires-predictions-sLd} (s$L_d$) and Fig.~\ref{fig-hires-predictions-Td} ($T_d$).

Having resolved maps of dust properties provides some important benefits over the global estimates of \citetalias{dobbels20}. First of all, even when one is only interested in the global properties of a galaxy, resolving the galaxy will lead to more accurate estimates. After all, a global SED leads to light-weighted averages, while we are usually more interested in mass-weighted averages \citep{utomo19}. A galaxy might contain a large amount of cold dust, but this dust might be outshined by warmer dust when grouped together on a global scale. This is similar to the outshining of old stars by young stars \citep{sorba15, sorba18, salim16}. Instead, if we resolve the galaxy, the warm and cold dust might be in separate locations, and this means that we can now detect the cold dust that was previously invisible. As a result, the dust mass derived from a global SED might be underestimated by about 50\%, and the dust temperature might be overestimated by a few degrees \citep{galliano11, galametz12, utomo19}. The better our resolution, the more we avoid this bias.

In addition, having a resolved dust map can also help to correct for dust attenuation. Radiative transfer simulations show that assuming a simple dust screen is often not appropriate \citep[e.g.][]{steinacker13, viaene17, williams19}. By resolving the galaxy, and estimating how much dust is present in each pixel, we can refine the correction for dust. Again, even when only interested in the global unattenuated SED, having a high resolution image helps. After all, different pixels can have different stellar populations, some of which will be more attenuated than others. Only by correcting for dust attenuation on a resolved scale can we properly take this into account.

Finally, resolved dust maps can be useful to validate numerical simulations of galaxy formation. These simulations have drastically improved over the past decade, and they provide a good match with observations \citep[e.g.][]{schaye15, furlong15, trayford15, springel18, nelson18, torrey19, ma18}. Some of the more recent simulations also follow the formation and destruction of dust grains \citep{aoyama18, dave19, granato20}. With many of the global scaling relations well reproduced, the goal of the next-generation simulations should be to reproduce resolved scaling relations. In this work, we showed that the dust properties can be predicted from UV--MIR fluxes on a resolved scale; these short wavelength fluxes are thus strongly related to the dust properties. This connection might be partially due to the dust attenuation, which affects the UV--NIR broadbands, but can also be due to the intrinsic stellar SED as well as the hot dust in the MIR. We can apply our model, trained on observations, and test it on simulated galaxies. Using radiative transfer, we can create synthetic observations of simulated galaxies \citep{camps18, parsotan21, kapoor21}. If the quality of those dust property predictions is close to that of observations, it is a sign that the simulations properly reproduce the star--dust relation of the real universe.

\section{Conclusions}
\label{sec-conclusions}

In this work we have predicted resolved maps of three dust properties, using only the UV to MIR broadband fluxes as input. These three properties are the specific dust luminosity (s$L_d$), specific dust mass (s$M_d$) and dust temperature ($T_d$). We use a random forest model, and train these on the DustPedia images at various pixel scales, ranging from 150 pc to 3 kpc. To avoid large galaxies from dominating the results, we weight each pixel inversely proportional to the number of pixels in that galaxy; this ensures that each galaxy---not each pixel---contributes equally. We find the following conclusions:

\begin{itemize}
	\item The dust luminosity can be predicted even better on an 18\arcsec resolution than on global scales, with an RMSE of 0.11 dex (compared to 0.16 dex on global scales). This is surprising, since energy balance is a better approximation at larger scales.
	\item The dust mass (RMSE = 0.32 dex) and temperature (RMSE = 3.15 K) can be predicted on 18\arcsec scales with an accuracy similar to that on global scales.
	\item While at first sight smaller scales seem to have worse predictions (see Fig.~\ref{fig-truevspred-sMd}), this difference is no longer significant when comparing the results on the same set of galaxies.
	\item About two-thirds of the RMSE can be explained as scatter between galaxies, rather than within galaxies. Pixels within a galaxy can be strongly correlated, and as a result predictions for individual galaxies can be biased. Typically, when s$M_d$ is overpredicted, $T_d$ will be underpredicted, and vice-versa; s$L_d$ is typically predicted very well.
	\item As shown in \citetalias{dobbels20}, we can estimate uncertainties for individual predictions, and these were verified in Fig.~\ref{fig-uncval}.
	\item We find M31 to be an outlier: its dust temperature is systematically overestimated, especially between the spiral arms. The dust luminosity is predicted very well, and thus the dust temperature is systematically underestimated. M33 is less biased, although we do not recover the temperature profile (decreasing temperature with increasing radius).
	\item Although we do use the pixel size (in pc) as a model input, it does not seem to be important. In other words, the relation between the starlight and the dust properties seems to have little dependence on the pixel size.
\end{itemize}

Our pipeline can be used to create FIR dust maps, with a higher resolution than \textit{Herschel}. These resolved dust maps can be used to improve the global dust property estimates of \citetalias{dobbels20}. In addition, they can be used to correct UV to NIR observations for dust attenuation. This then leads to improved estimates of stellar properties. Finally, they can be used as a test for cosmological simulations. Recent simulations can reproduce many global scaling relations well, and so the next step should be to verify resolved scaling relations. The model that we propose is essentially a scaling relation between UV--MIR broadbands and the dust properties. In the coming years, excellent UV--MIR telescopes will become active, but there is no FIR telescope confirmed as of now. With the help of our model, we can use these upcoming telescopes to infer useful information about the dust. 

\begin{acknowledgements}
	The authors gratefully acknowledge support from the Flemish Fund for Scientific Research (FWO-Vlaanderen). W.D. is a pre-doctoral researcher of the FWO-Vlaanderen (application number 1122718N). Special thanks to the Flemish Supercomputer Centre (VSC) for providing computational resources and support. We gratefully acknowledge the support of NVIDIA Corporation with the donation of the Titan Xp GPU used for this research. Following open-source python projects were used extensively: scikit-learn \citep{pedregosa12}, CIGALE, numpy, scipy, pandas, matplotlib. Many thanks to the maintainers and contributors of these projects. 
	
	DustPedia is a collaborative focused research project supported by the European Union under the Seventh Framework Programme (2007-2013) call (proposal no. 606824). The participating institutions are: Cardiff University, UK; National Observatory of Athens, Greece; Ghent University, Belgium; Université Paris Sud, France; National Institute for Astrophysics, Italy and CEA (Paris), France.
	
	Thank you to Angelos Nersesian, Ana Tr\v{c}ka, and Viviana Acquaviva for the valuable feedback throughout this work. The authors thank the anonymous referee for a prompt and constructive referee report.
\end{acknowledgements}

\bibliographystyle{aa} 
\bibliography{references} 

\begin{appendix}
	
\section{Additional figures}
\label{app-additional-figures}

\begin{figure*}
	\centering
	\includegraphics[width=17cm]{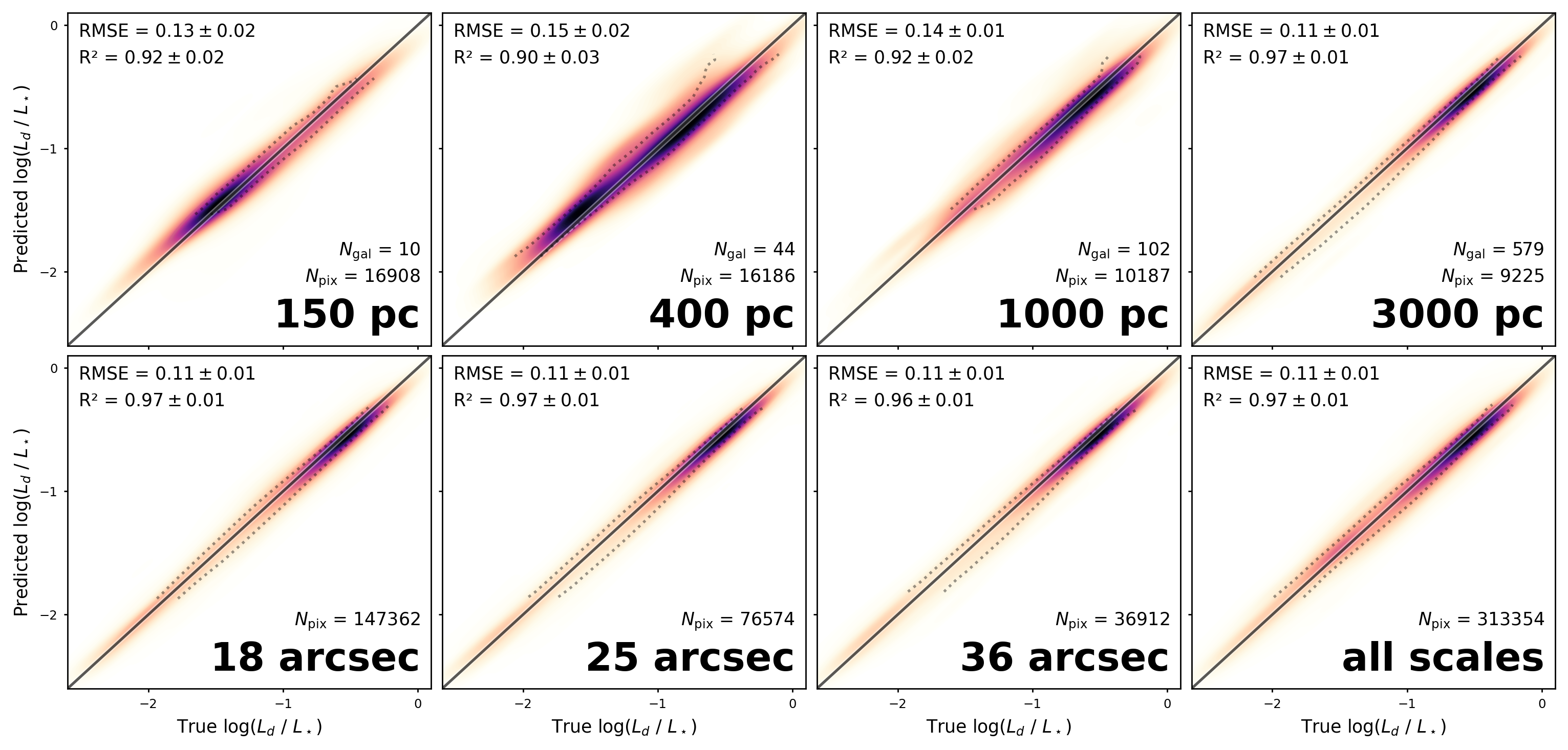}
	\caption{Similar to Fig.~\ref{fig-truevspred-sMd}, but for the specific dust luminosity.}
	\label{fig-truevspred-sLd}
\end{figure*}

\begin{figure*}
	\centering
	\includegraphics[width=17cm]{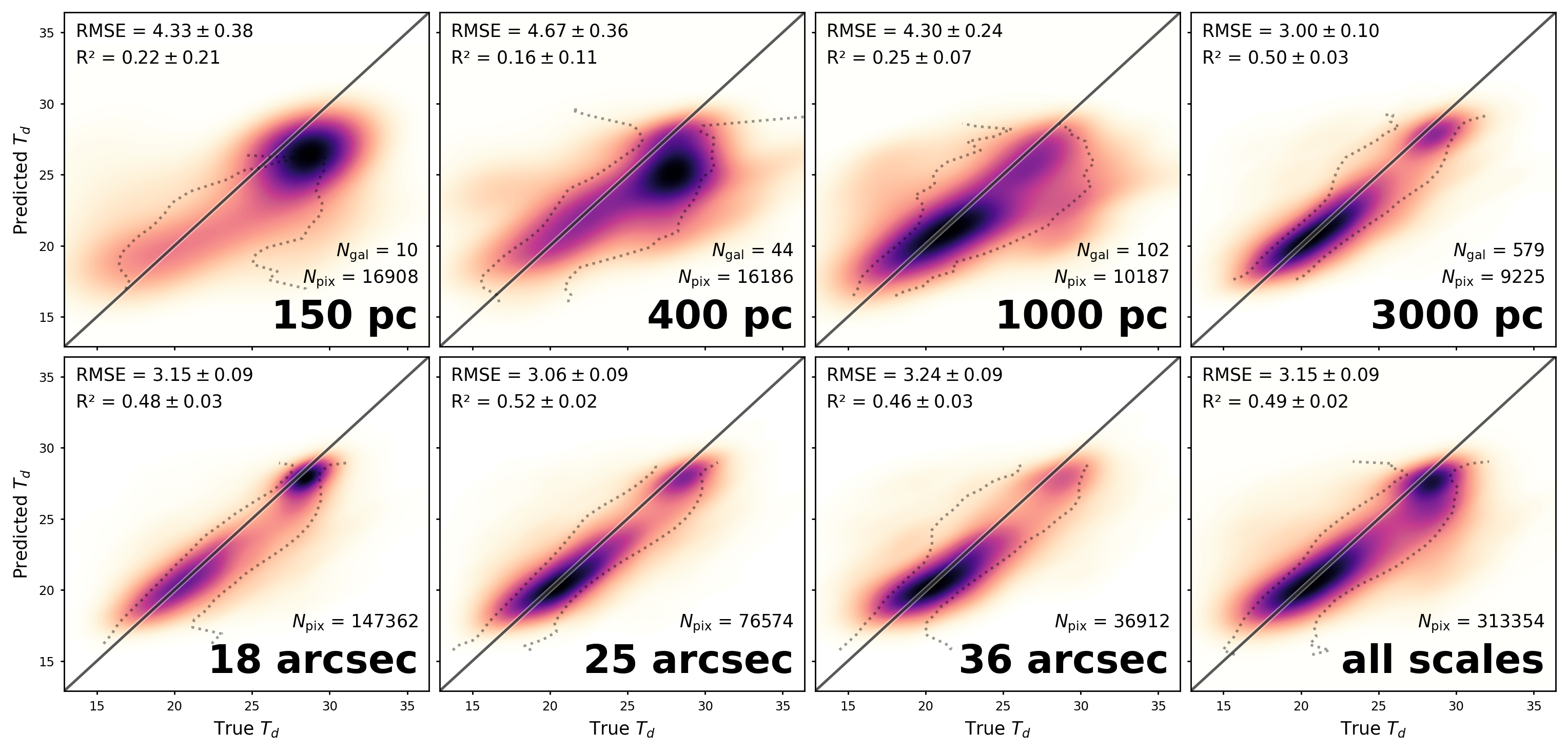}
	\caption{Similar to Fig.~\ref{fig-truevspred-sMd}, but for the dust temperature.}
	\label{fig-truevspred-Td}
\end{figure*}

This appendix includes additional figures similar to those in the main text. Fig.~\ref{fig-truevspred-sLd} compares the predictions and ground truth for the specific dust luminosity at various scales, and is similar to Fig.~\ref{fig-truevspred-sMd} in the main text. Fig.~\ref{fig-truevspred-Td} does the same, but for the dust temperature. These figures are described shortly in Sect.~\ref{sect-crossscale}.

\begin{figure*}
	\centering
	\includegraphics[width=17cm]{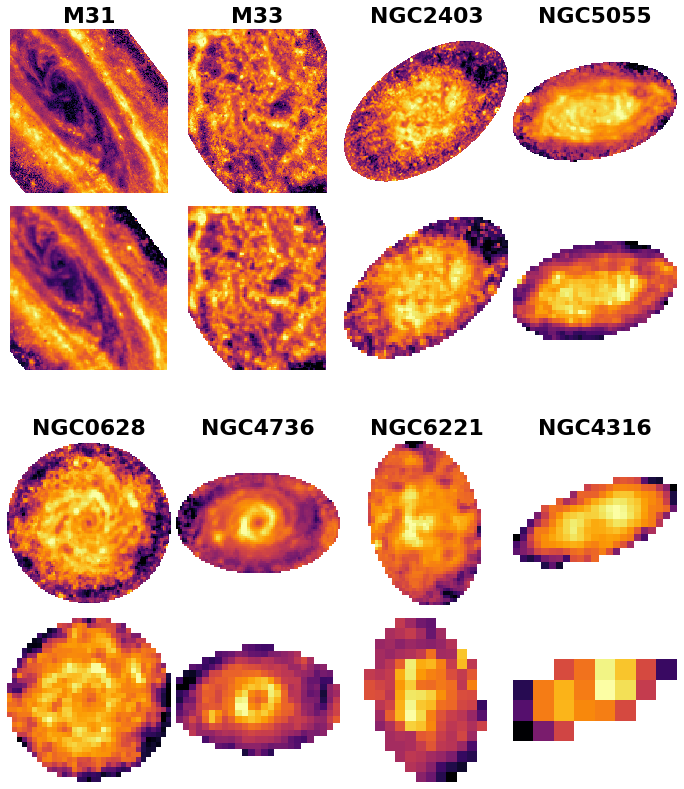}
	\caption{Similar to Fig.~\ref{fig-hires-predictions-sMd}, but for the specific dust luminosity.}
	\label{fig-hires-predictions-sLd}
\end{figure*}

\begin{figure*}
	\centering
	\includegraphics[width=17cm]{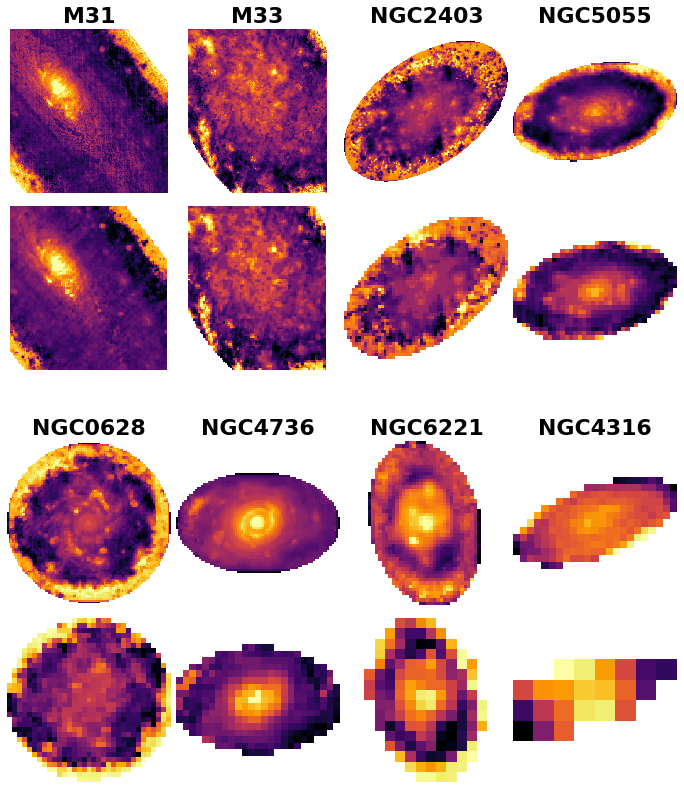}
	\caption{Similar to Fig.~\ref{fig-hires-predictions-sMd}, but for the specific dust temperature.}
	\label{fig-hires-predictions-Td}
\end{figure*}

Fig.~\ref{fig-hires-predictions-sLd} shows the high resolution predictions of the specific dust luminosity, similar to Fig.~\ref{fig-hires-predictions-sMd} in the main text. Fig.~\ref{fig-hires-predictions-Td} does the same for the dust temperature. These figures are shortly described in Sect.~\ref{ssec-highres-dustmap}. 

\end{appendix}
\end{document}